\documentclass[aps, onecolumn,floatfix, preprintnumbers,superscriptaddress, 11pt]{revtex4-2}

\usepackage{setspace}
\usepackage{graphics,amssymb,amsmath,epsfig,color}
\usepackage{graphicx}
\usepackage{graphicx, color, xcolor, soul}

\usepackage[raggedright, small]{titlesec}

\titlespacing*{\section}
  {0pt}{2.0ex plus 1ex minus .2ex}{0.8ex plus .2ex} 

\titlespacing*{\subsection}
  {0pt}{1.5ex plus 1ex minus .2ex}{0.6ex plus .2ex}
\makeatletter

\def\@hangfrom@section#1#2#3{\@hangfrom{#1#2}#3}
\def\@hangfroms@section#1#2{#1#2}

\begin{document}

\title{
\Large{Chip-scale modulation-free laser stabilization using vacuum-gap micro-Fabry-P\'erot cavity}
}

\author{Mohamad Hossein Idjadi}
\email{Corresponding author: mohamad.idjadi@nokia-bell-labs.com}
\affiliation{Nokia Bell Labs, 600 Mountain Ave., Murray Hill, NJ 07974, USA.}

\author{Haotian Cheng}
\affiliation{Department of Applied Physics, Yale University, New Haven, CT 06511, USA}

\author{Farshid Ashtiani}
\affiliation{Nokia Bell Labs, 600 Mountain Ave., Murray Hill, NJ 07974, USA.}

\author{Benjia Li}
\affiliation{Department of Applied Physics, Yale University, New Haven, CT 06511, USA}

\author{Kwangwoong Kim}
\affiliation{Nokia Bell Labs, 600 Mountain Ave., Murray Hill, NJ 07974, USA.}

\author{Naijun Jin}
\affiliation{Department of Applied Physics, Yale University, New Haven, CT 06511, USA}

\author{Franklyn Quinlan}
\affiliation{Time and Frequency Division, National Institute of Standards and Technology, Boulder, CO 80305, USA}

\author{Peter T. Rakich}
\email{Corresponding author: peter.rakich@yale.edu}
\affiliation{Department of Applied Physics, Yale University, New Haven, CT 06511, USA}

\begin{abstract}
Narrow-linewidth lasers are vital for a broad range of scientific and technological applications, including atomic clocks and precision sensing. Achieving high frequency stability is often as critical as ensuring scalability, portability, and cost-effectiveness in the development of low noise laser systems. Conventional electro-optic stabilization techniques, such as Pound–Drever–Hall locking to ultra-high-finesse resonators held in a vacuum chamber, provide excellent performance but remain challenging to scale. 
Here, we propose and experimentally demonstrate a cavity-coupled interferometric laser stabilization technique implemented on a silicon photonic chip and integrated with a compact, scalable micro-Fabry-P\'erot cavity. The vacuum-gap optical cavity operates in air, achieving a quality factor of approximately $2.0\times10^9$ and a fractional frequency instability of $1.45\times10^{-12}$ at one-second averaging time. Integration of the proposed technique with the compact cavity yields more than 38-fold reduction in the laser’s integrated linewidth and nearly three orders of magnitude suppression of frequency noise at 10 Hz offset frequency. The hybrid-integration of the proposed photonic chip with the micro-Fabry-P\'erot cavity establishes a scalable and portable route toward chip-integrated ultra-stable lasers, paving the way for precision optical systems deployable beyond laboratory environments.
\end{abstract}

\maketitle

\section*{Introduction}
Stable lasers underpin advances in metrology \cite{udem2002optical}, sensing \cite{marra2018ultrastable, xu2019sensing}, spectroscopy \cite{demtroder1973laser}, optical atomic clocks \cite{ludlow2015optical, jiang2011making}, and photonic microwave generation \cite{li2013microwave, kudelin2023photonic, sun2023integrated, he2024chip}. 
% ...
In such applications, where narrow-linewidth lasers play a vital role, the requirements extend beyond achieving both long- and short-term frequency stability. Equally important is the development of solutions that are low-cost, scalable, and compact, while maintaining low phase noise performance. These factors are critical for unlocking the full potential and enabling the widespread deployment of low noise laser technology. In other words, not only must the components and building blocks of a stable laser system be scalable and miniature, but the method used to suppress laser frequency noise must also be effective and simple enough to implement at scale and low cost. 

A variety of methods ranging from optical \cite{liang2015ultralow, dahmani1987frequency, cheng2025harnessing} and electro-optic feedback \cite{drever1983laser, matei20171, idjadi2020nanophotonic} to feed-forward \cite{chen2019feedforward, aflatouni2012wideband} techniques have been developed to suppress laser frequency noise and improve long- and short-term and stability. Electro-optic (EO) feedback stabilization leverages mature and robust electronic-photonic techniques to detect laser frequency noise in the optical domain and process it electronically. The resulting error signal is used to control the laser frequency actively, enabling tight locking to an optical reference and ensuring long-term stability.  In general, EO stabilization systems comprise two key components: an EO circuit that detects and processes the error signal, and a stable optical resonator or cavity that serves as the frequency reference to which the laser frequency is compared. The Pound–Drever–Hall (PDH) technique \cite{drever1983laser, black2001introduction} is one of the most established EO methods for suppressing laser phase noise. In a PDH feedback loop, the laser is phase-modulated by an external EO modulator, filtered by an optical resonator, and then photodetected to generate the error signal used for controlling the laser frequency. 
To achieve monolithic integration of a PDH system on a chip and realize integrated and scalable low-noise lasers, the photonic platform must simultaneously support efficient, low-loss EO phase modulation, and ultra-low-loss waveguides capable of realizing high quality-factor (Q-factor) resonators and delivering a high signal-to-noise ratio (SNR) to the in-loop electronics—a combination that remains challenging to realize within a single material system.

For instance, silicon remains the most attractive integrated photonic platform owing to its device reliability, scalability, broad component library, and compatibility with monolithic complementary metal-oxide-semiconductor (CMOS) integration \cite{idjadi2017integrated, siew2021review, xiang2021perspective}. However, the propagation loss of silicon nanophotonic waveguides exceeds that of silicon nitride (SiN) by nearly two orders of magnitude, fundamentally limiting the achievable Q-factors in silicon-based resonators and limiting PDH loop performance. In contrast, SiN offers ultra-low-loss waveguides, with propagation losses below 0.4 dB m$^{-1}$, enabling high Q-factor integrated resonators and high PDH error signal SNR \cite{cheng2025spiral, liu202236}. Nevertheless, despite the excellent performance of partially integrated PDH systems and recent progress in integrating EO modulators on SiN \cite{wang2022silicon}, large-scale monolithic demonstrations of such modulators in commercially available foundries remain challenging.

To address this trade-off, we recently demonstrated the modulation-free cavity-coupled Mach-Zehnder interferometer (CCMZI) laser stabilization technique that enables the realization of narrow-linewdith lasers on arbitrary integrated photonic platforms \cite{idjadi2024modulation, idjadi2025modulation_optexp}. Here, “modulation-free” specifically indicates that no EO phase modulation of the laser is required to generate the error signal. In this interferometric method, one arm of the interferometer is coupled to an optical resonator, and its output interferes with a reference path on balanced photodetectors (BPDs). The resulting electronic error signal, used to control the laser frequency, is thus generated without requiring EO phase modulation of the laser. This method is versatile and not limited to a specific optical resonator (e.g. integrated ring resonator or Fabry–P\'erot cavity) or material platform. The modulation-free CCMZI architecture therefore eliminates unwanted residual amplitude modulation associated with EO phase modulation in the PDH loop \cite{shi2018suppression} and enables a simpler electronic architecture for error signal processing in the feedback loop. 

As mentioned earlier, an optical frequency reference is a fundamental element of any laser stabilization system, irrespective of the stabilization method employed. Numerous high Q-factor optical resonators and cavity platforms have been demonstrated, which can be broadly categorized into dielectric resonators \cite{ji2021methods, puckett2021422, ye2022integrated, zhang2021ultra} and vacuum-gap cavities \cite{kelleher2023compact, davila2017compact}. Dielectric resonators provide the advantage of on-chip integration whereas vacuum-gap cavities, although bulkier and more difficult to engineer for compactness, scalability, and portability, can achieve more than three orders of magnitude lower fractional frequency instability at one-second averaging times \cite{liu202236, zhang2019microrod, mclemore2022miniaturizing, liu2024ultrastable}. To overcome this trade-off, we recently demonstrated a wafer-scale, compact vacuum-gap micro-Fabry–P\'erot ($\mu$FP) cavity with a volume below 0.6 cm$^3$ and a Q-factor exceeding $10^{9}$ in the C-band, delivering excellent frequency stability while operated in air \cite{jin2022micro, cheng2025harnessing, liu2024ultrastable}.

Here, we report a compact and portable laser stabilization system operating at atmospheric pressure and room temperature, which combines, for the first time, an integrated modulation-free silicon photonic CCMZI chip with a miniature vacuum-gap ultra-high Q-factor $\mu$FP cavity that achieves sub-Hz fundamental linewidth sensitivity. The system suppresses frequency noise of a laser by nearly three orders of magnitude at a 10 Hz offset, reduces the integrated linewidth by more than 38-fold, and achieves a one-second fractional frequency instability of $1.45\times10^{-12}$, more than an order of magnitude lower than that of the state-of-the-art SiN integrated coil resonator \cite{liu202236}, while remaining scalable to wafer-level fabrication and compatible with monolithic CMOS electronic integration.

\section*{Results}

\subsection*{Principle of operation}
\begin{figure}[tb]
    \centering
    \includegraphics[width=\textwidth]{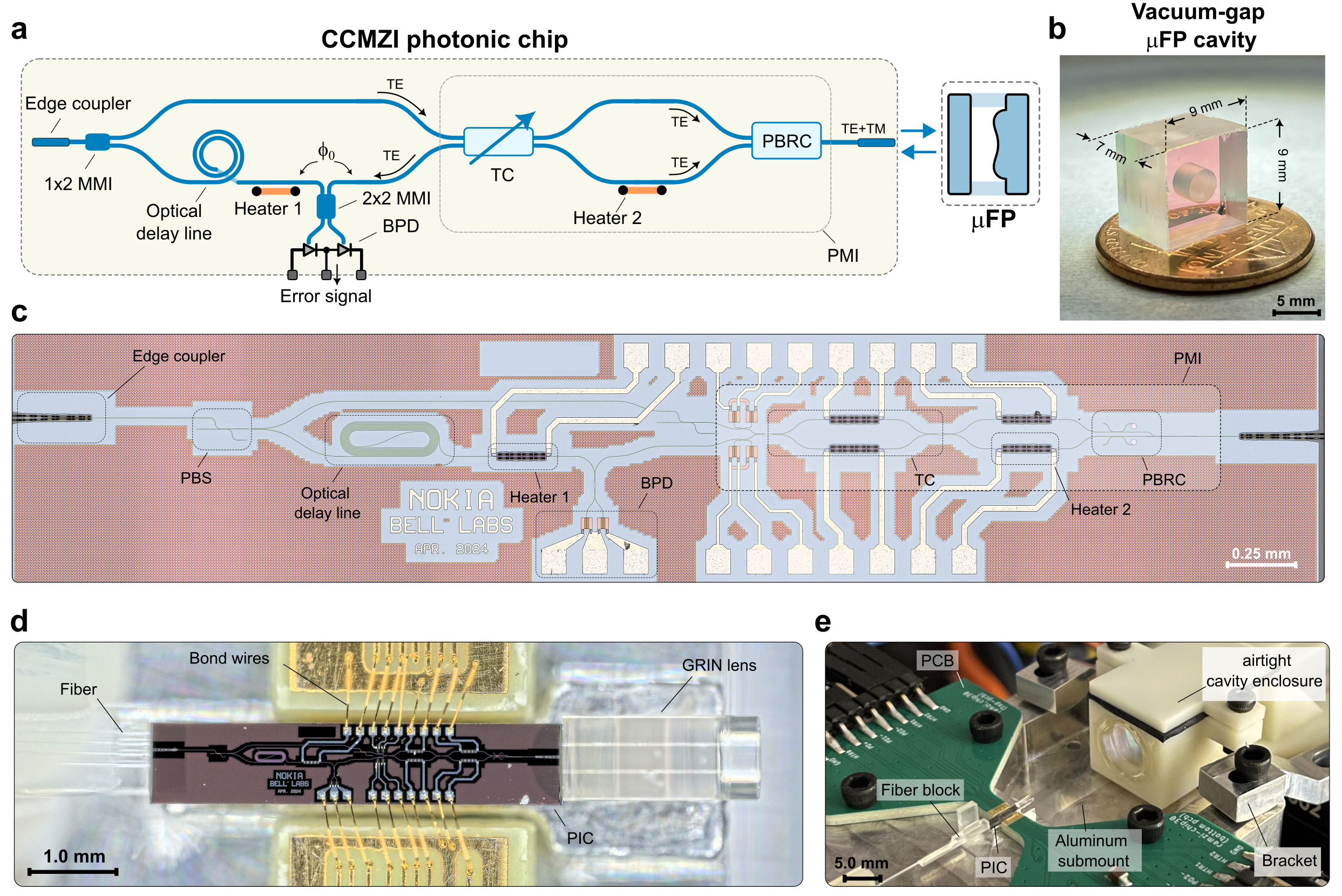}
    \caption{\textbf{Cavity-coupled MZI (CCMZI) laser stabilization.} \textbf{a,} Block diagram of the CCMZI photonic chip. \textbf{b,} Photograph of the vacuum-gap micro-Fabry–P\'erot ($\mu$FP) cavity. The ultra-low-expansion (ULE) spacers have a thickness of 3 mm, corresponding to a free spectral range of 50 GHz. The diced cavity dimensions are 9 mm $\times$ 9 mm $\times$ 7 mm. \textbf{c,} Microphotograph of the implemented CCMZI photonic integrated circuit (PIC), fabricated using the Advanced Micro Foundry (AMF) standard silicon-on-insulator (SOI) photonic process. \textbf{d,} Packaged PIC with a single-mode fiber at its input and a gradient-index (GRIN) lens at the output. \textbf{e,} The packaged PIC is mounted on a temperature-controlled aluminum submount, with output light coupled to the cavity through free-space. The $\mu$FP cavity is enclosed in a custom-designed, 3D-printed airtight housing for enhanced acoustic shielding. Details of PIC and cavity packaging are discussed in Methods. MMI multi-mode interferometer, BPD balanced photodetector, TC tunable coupler, PBRC polarization beam rotator-combiner, PBS polarization beam splitter, PMI poor man’s isolator.}
    \label{fig_pic}
\end{figure}
Figure \ref{fig_pic}a shows the block diagram of the CCMZI laser stabilization photonic chip. The laser is coupled into the fundamental TE mode of the nanophotonic waveguide via an edge coupler (EC). The input light is then split into two paths by a 1$\times$2 multi-mode interferometer (MMI) with 50\% splitting ratio. The use of an MMI ensures broadband operation and reduces sensitivity to fabrication-induced errors. The waveguide carrying the TE mode feeds the Poor Man’s Isolator (PMI). The on-chip PMI, developed based on our previous work \cite{cheng2023novel}, consists of a tunable coupler (TC) that splits the input signal into the desired ratio at its outputs. These two outputs are directed into a polarization beam rotator-combiner (PBRC), where one of the TE inputs is converted to TM polarization and then combined with the remaining TE input. Both TE and TM modes are launched toward the $\mu$FP cavity through an output EC. A gradient refractive index (GRIN) lens is utilized to convert the mode profile of the EC to match that of the $\mu$FP cavity. The cavity-reflected modes are converted back to TE polarization and recombined at the tunable coupler. As shown in Fig. \ref{fig_pic}a, Heater 2 adjusts the relative phase between the two modes to ensure maximum reflection cancellation. It is worth noting that a 2$\times$2 MMI or symmetric coupler would suffice ideally; however, the tunable coupler compensates for loss imbalance between the two modes and thereby enables maximum suppression of the reflected field. Additional details of PMI operation are provided in Supplementary Note 1. 

On the lower branch, an optical spiral delay line compensates the free-space delay between the PIC and the $\mu$FP cavity, thereby suppressing unwanted interference fringes. The optical phase of the delayed signal is adjusted by a thermal phase shifter (Heater 1 in Fig. \ref{fig_pic}a) to ensure proper interference conditions within the MZI. This branch then interferes with the output of the PMI in a 2$\times$2 MMI, and the resulting signals are detected by BPD to generate the error signal for laser locking. 
Generally, if the transfer function of the optical frequency reference (e.g. transmission of a microring resonator or reflection from the FP cavity) is denoted as $T(\omega)$, then the corresponding error signal generated by the CCMZI chip can be expressed as \cite{idjadi2024modulation}
\begin{equation}
    i_{error}(\omega)=RP_0|T(\omega)|\ \mathrm{sin}(\angle T(\omega)-\phi_0), \label{eq1_error}
\end{equation}
where $\omega, |T(\omega)|,\angle T(\omega), \phi_0, P_0$, and $R$ are laser frequency, the amplitude and phase of the optical reference transfer function at the frequency of $\omega$, phase difference between the arms of the MZI controlled by Heater 1, optical power of at the input of the MZI, and the responsivity of the photodetectors (PDs), respectively. As described by Eq. (\ref{eq1_error}), appropriate adjustment of the constant phase $\phi_0$, via a thermal phase shifter, ensures an asymmetric error signal about the frequency reference, which can then be used for laser frequency locking.  Details of the error signal derivation is discussed in Supplementary Note 2.
\subsection*{A compact, portable, and stable optical frequency reference}
As mentioned earlier, optical resonators are a key component, serving as a frequency reference, in optical and electro-optic laser stabilization systems which can be broadly categorized as dielectric, typically integrated on a chip, or vacuum-gap cavities. For a given mode volume and operating temperature, vacuum-gap resonators generally provide superior low noise performance. The primary challenge, however, is to realize a portable, miniature, ultra-stable cavity that can maintain a vacuum gap between its mirrors while operating under atmospheric pressure. Figure \ref{fig_pic}b shows a photograph of the developed vacuum-gap $\mu$FP cavity, which achieves an ultra-high Q-factor of $2.0\times10^9$ while remaining compact and portable. The $\mu$FP cavity comprises a concave mirror with a 35~cm radius of curvature \cite{jin2022micro}, a planar mirror, and a 3~mm-thick spacer made of ultra-low-expansion (ULE) glass to suppress temperature-induced drift~\cite{cheng2025harnessing}. This configuration supports a fundamental optical mode with a $240~\mathrm{\mu m}$ mode-field diameter. The entire structure is wafer-level bonded in vacuum and subsequently singulated while maintaining a hermetically sealed vacuum gap inside, ensuring long-term vacuum integrity without the need for additional vacuum enclosures. Details of the cavity ring-down experiment are provided in Supplementary Note 3.
\subsection*{CCMZI photonic chip}
Figure \ref{fig_pic}c shows the implemented CCMZI photonic chip, fabricated using the commercially available Advanced Micro Foundry (AMF) silicon photonic process, with a total footprint of 4.0 mm$^2$. As shown in Fig. \ref{fig_pic}d, the PIC is mounted on an aluminum submount and packaged with an optical fiber at the input and a GRIN lens at the output for mode size matching. The $\mu$FP cavity is enclosed in a 3D-printed airtight housing and aligned with the packaged PIC, as illustrated in Fig. \ref{fig_pic}e. The airtight enclosure isolates the cavity from acoustic noise, which would otherwise couple back into the laser as phase noise. Further details of the photonic chip implementation and the PIC–$\mu$FP cavity packaging are provided in Methods.

\subsection*{Open loop characterization}
\begin{figure}[tb]
    \centering
    \includegraphics[width=\textwidth]{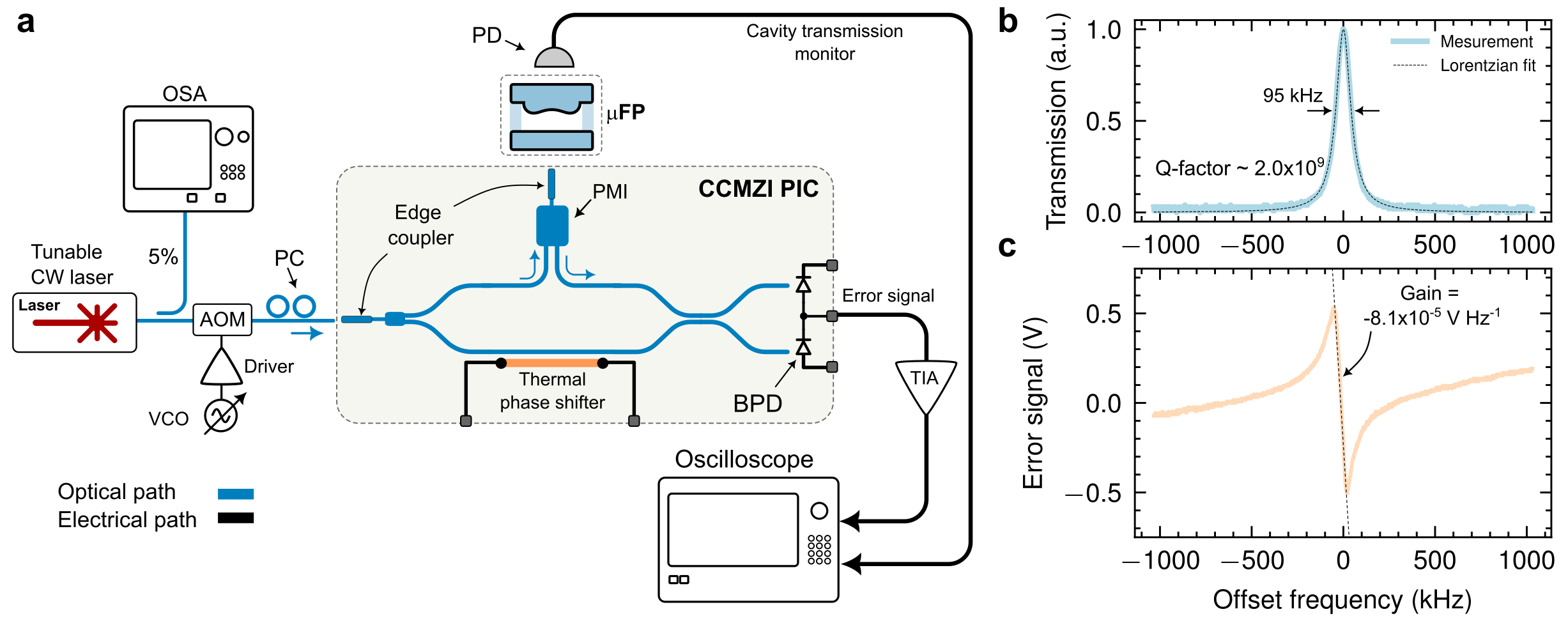}
    \caption{\textbf{Open-loop characterization.} \textbf{a,} Measurement setup for CCMZI open-loop response and $\mu$FP cavity characterization. \textbf{b,} Measured transmission of the $\mu$FP cavity, showing a resonance linewidth of 95 kHz, corresponding to a Q-factor of $2.0\times 10^9$ at wavelength of 1550.218 nm. \textbf{c,} Measured error signal of the CCMZI circuit. PC, polarization controller; VCO, voltage-controlled oscillator; OSA, optical spectrum analyzer; PD, photodetector; TIA, transimpedance amplifier. }
    \label{fig_openloop}
\end{figure}
Figure \ref{fig_openloop}a shows the measurement setup used to characterize the open-loop CCMZI response and the $\mu$FP cavity resonance. A tunable continuous-wave (CW) fiber laser at wavelength of 1550.218 nm is coupled into an acousto-optic modulator (AOM) that serves as a frequency shifter, due to the lack of fast frequency modulation port on the laser, driven by a voltage-controlled oscillator (VCO). By sweeping the VCO, the laser frequency is scanned across the cavity resonance. The cavity transmission is monitored by a PD, while the PIC output current is amplified and converted to a voltage signal (i.e. the error signal) using a transimpedance amplifier (TIA), and then recorded on an oscilloscope.\\
Figure \ref{fig_openloop}b shows the $\mu$FP cavity resonance, where the measured full-width at half-maximum (FWHM) linewidth is approximately 95 kHz that matches the measured photon decay time of about 1.6 $\mu$sec shown in Supplementary Fig. 3b. The free-spectral range of the cavity is 50 GHz and the corresponding Q-factor and finesse are about $2.0\times10^9$ and $536\times10^3$, respectively. Figure \ref{fig_openloop}c shows the measured error signal with asymmetric response centered about the $\mu$FP resonance frequency which matches the theoretical calculation discussed in Supplementary Note 2.   

\subsection*{Laser frequency noise suppression}
\begin{figure}[htb]
    \centering
    \includegraphics[width=\textwidth]{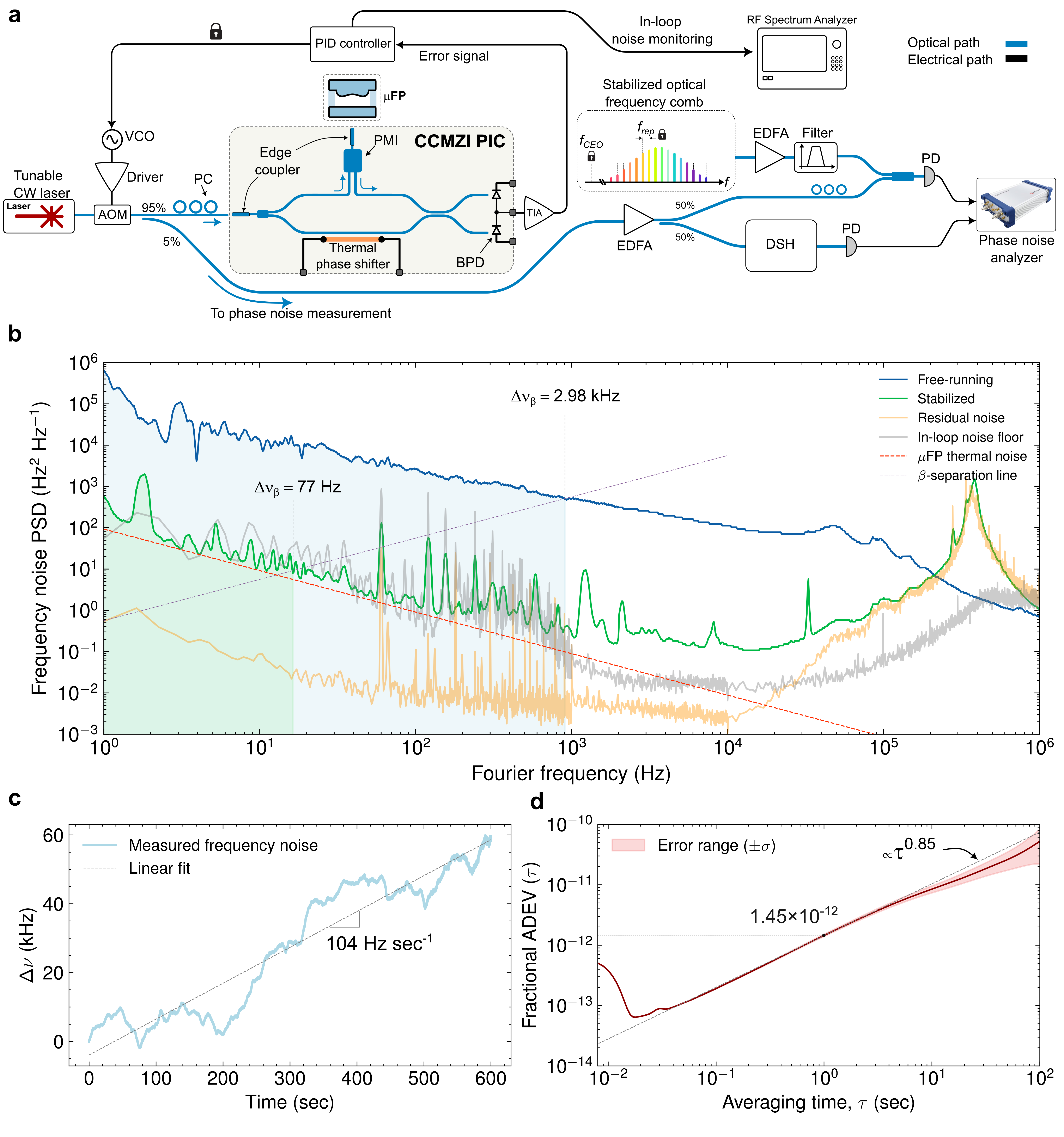}
    \caption{\textbf{The closed-loop operation.} \textbf{a,} Measurement setup for laser frequency stabilization and phase noise measurement. \textbf{b,} Power spectral density (PSD) of the free-running and stabilized laser. The integrated linewidth, $\Delta\nu_{\beta}$, is obtained by averaging the frequency noise PSD from 1 Hz up to the frequency at which the PSD intersects the $\beta$-separation line \cite{di2010simple}. \textbf{c,} Stabilized laser beat-note fluctuations measured over 10 minutes. \textbf{d,} Modified Allan deviation of the fractional frequency instability when the laser is locked to the $\mu$FP cavity. Further details are provided in Methods.}
    \label{fig_closedloop}
\end{figure}

Figure \ref{fig_closedloop}a shows the block diagram of the closed-loop operation and laser frequency noise suppression setup. The CW laser at wavelength of 1550.218 nm is frequency-shifted by the AOM and coupled into the on-chip waveguide using a single-mode fiber glued to the PIC. The thermal phase shifters (Heaters 1 and 2 in Fig. \ref{fig_pic}a) are adjusted to minimize reflections via the PMI and to maximize the error signal amplitude. The PIC output is amplified by a TIA, with 10 k$\Omega$ of transimpedance gain, and applied to the input of the PID controller. The PID output drives the control voltage of the VCO, which in turn shifts the laser frequency and locks it to the resonance frequency of the $\mu$FP cavity. The error signal is also monitored on a RF spectrum analyzer for in-loop residual noise monitoring.
As shown in Fig. \ref{fig_closedloop}a, a small portion of the laser output is used for phase noise measurement and frequency counting. Half of the tapped-out power is mixed with the output of a fully-stabilized optical frequency comb for beat note measurement and long-term stability test (Fourier frequency $<$ 1 kHz) and the rest of the power is used in a delayed-self heterodyne (DSH) setup for short-term stability characterization (Fourier frequency $>$ 1 kHz).  

Figure \ref{fig_closedloop}b shows the measured power spectral densities (PSDs) of frequency noise of the free-running and stabilized laser. The CCMZI stabilization system achieves frequency noise performance approaching the theoretical thermal noise limit of the $\mu$FP cavity, with about 30 dB frequency noise suppression at 10 Hz offset and, as indicated by the in-loop noise floor, sub-hertz fundamental linewidth sensitivity. The integrated linewidth ($\Delta\nu_{\beta}$), as depicted in Fig. \ref{fig_closedloop}b, is reduced from 2.98 kHz in the free-running case to 77 Hz when stabilized, representing more than a 38-fold improvement. Moreover, the residual in-loop noise PSD confirms that the laser remains tightly locked to the $\mu$FP cavity and the stabilized laser frequency noise is limited by the total in-loop noise of the system for offsets below 1 kHz.

Figures \ref{fig_closedloop}c and \ref{fig_closedloop}d show the long-term stability performance when the laser is locked to the cavity. Figures \ref{fig_closedloop}c and \ref{fig_closedloop}d present the frequency drift of the stabilized laser over a 10-minute interval and the corresponding modified Allan deviation (ADEV) of the fractional frequency instability. A $\tau^{0.85}$ dependence and a fractional frequency instability of $1.45\times10^{-12}$ at 1 sec averaging time are measured that indicate a linear drift of approximately 104 Hz sec$^{-1}$, attributed to slow thermal drift of the cavity rather than the PIC.  
\section*{Discussion}
Mechanical and thermal packaging are critical yet challenging aspects of ultra-low-noise laser design. Integrated photonic systems benefit from their small footprint, which reduces sensitivity to temperature gradients and acoustic vibrations. In contrast, although larger optical cavities achieve lower fundamental thermal noise, they are more susceptible to acoustic disturbances and temperature drift. This trade-off underscores the importance of proper packaging. 
\begin{figure}[tb]
    \centering
    \includegraphics[width=\textwidth]{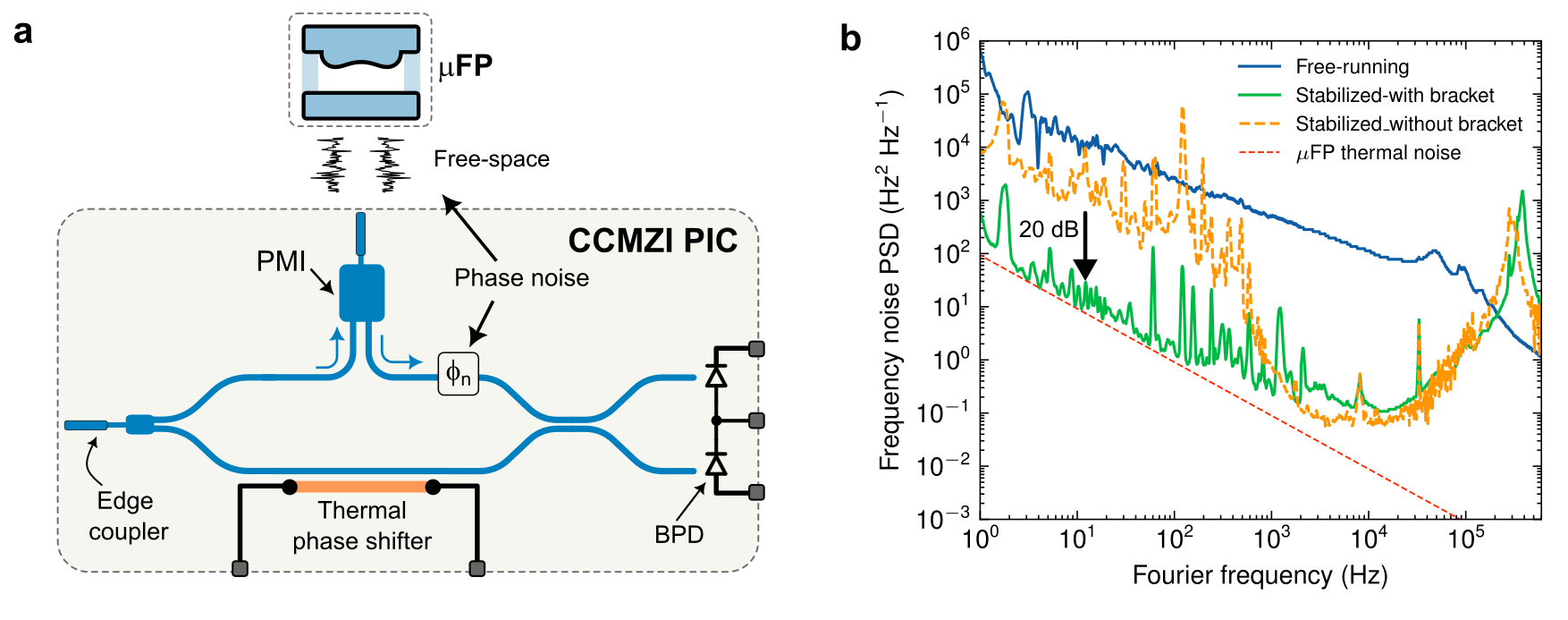}
    \caption{\textbf{Effect of mechanical vibration on laser phase noise.} \textbf{a,} Effect of mechanical instability in CCMZI architecture. \textbf{b,} Measured PSD of laser frequency noise locked to the $\mu$FP with and without the aluminum brackets. }
    \label{fig_bracket}
\end{figure}
Although ULE spacer is used between the mirrors, the $\mu$FP cavity is not temperature-stabilized at the ULE zero-crossing temperature, where the coefficient of thermal expansion passes through zero \cite{fox2009temperature}. Consequently, as shown in the experiment results in Fig. \ref{fig_closedloop}c and \ref{fig_closedloop}d, a linear drift of the $\mu$FP resonance frequency is observed. Improved packaging with an acoustic shield and active temperature stabilization at the zero-crossing point can greatly reduce the slow frequency drift \cite{kelleher2023compact, liu2024ultrastable}. Nevertheless, the demonstrated $\mu$FP cavity, even without active temperature control, achieves 27-fold reduction in fractional frequency instability at one-second averaging time and 26-fold reduction in frequency drift compared to state-of-the-art silicon nitride resonators \cite{liu202236}, highlighting the superiority of the proposed technology over integrated dielectric resonators.

Moreover, mechanical stability also strongly affects frequency noise performance, typically at Fourier frequencies below 1 kHz. We observed that placing the $\mu$FP cavity inside an airtight enclosure significantly reduces the stabilized laser’s phase noise. Details of the $\mu$FP cavity enclosure is disclosed in Methods. Particularly in the CCMZI architecture, any mechanical instability or displacement between the cavity and the PIC introduces excess phase noise. For example, as shown in Fig. \ref{fig_bracket}a, when the PIC and cavity holder are not clamped with an aluminum bracket (Fig. \ref{fig_pic}e), the resulting mechanically induced round-trip phase noise behaves like an additional phase term in the MZI. This phase noise is converted into intensity fluctuations in the electronic domain, analogous to intrinsic laser phase noise, and substantially degrades the closed-loop noise performance. Figure \ref{fig_bracket}b shows that the stabilized laser’s frequency noise PSD improves by more than 20 dB when the cavity holder is simply clamped to the PIC submount. In future designs, positioning the cavity in close proximity to the PIC and bonding it to the submount inside a temperature-controlled, airtight enclosure will enable more robust, low noise operation and further reduces the overall form factor.  

It is worth emphasizing that we employed an AOM as a frequency shifter because the low-noise CW fiber laser used in the experiments lacked a frequency-modulation port with sufficient bandwidth (limited to 20 kHz). AOMs typically introduce 3-4 dB of optical insertion loss and a latency on the order of a few hundred nanoseconds, determined by the acoustic velocity and device geometry. As shown in Fig. \ref{fig_closedloop}b, the loop delay, including AOM latency, limits the maximum stable loop gain and produces a servo bump around 400 kHz. Frequency shifting with an AOM also requires a VCO and a power amplifier to efficiently drive the AOM, further contributing to latency and overall power consumption. Alternatively, commercial low-noise lasers, with comparable free-running phase noise, equipped with high-speed direct frequency-modulation ports (up to 10 MHz) \cite{indieLXM} can eliminate the need for an AOM, thereby reducing optical loss and improving closed-loop noise performance.

In conclusion, we have demonstrated, for the first time, a portable and compact modulation-free laser frequency stabilization scheme based on a CCMZI combined with a vacuum-gap $\mu$FP cavity operating at atmospheric pressure. This approach achieves a sub-Hz fundamental linewidth sensitivity. When applied to a low noise fiber laser, the architecture yields more than a 38-fold reduction in the integrated linewidth and nearly 30 dB frequency noise suppression at 10 Hz offset. The PIC, implemented on the commercially available AMF SOI platform, occupies only 4.0 mm$^2$ and consumes total of 3.45 mW of power for the thermal phase shifters. The vacuum-gap $\mu$FP cavity, with a volume of less than 0.6 cm$^3$, delivers unprecedented phase noise performance while remaining compact and portable. The co-integration of the PIC and $\mu$FP cavity within this modulation-free CCMZI framework paves the way toward scalable, low-cost, and portable low noise lasers and optical frequency references for a wide range of scientific and technological applications.  
\section*{Methods}
\subsection*{Silicon photonic chip}
All photonic devices are monolithically integrated in AMF commercially available standard SOI photonic process. The laser is coupled into the chip using an on-chip optical EC with approximately 3 dB insertion loss. The nominal mode size of the edge coupler is 9 $\mu$m $\times$ 5 $\mu$m. The input light is passed through a polarization beam splitter (PBS), with nominal insertion loss of 0.2 dB, to ensure operation in TE mode. Then the signal is split in half using a broadband 1$\times$2 MMI with 50\% splitting ratio and nominal excess loss of 0.1 dB. The light traveling in the top branch of the MZI enters the on-chip PMI which provides more than 10 dB of isolation. Details of the PMI operation is discussed in Supplementary Note 1. The output of the PMI utilizes a PBRC with nominal TE-TE and TE-TM conversion insertion loss of about 0.4 dB and 0.25 dB, respectively. The thermo-optic phase shifters used in the PMI and MZI, surrounded by thermal isolation trenches, exhibit a $P_{\pi}$ of 1.36 mW. Two pairs of germanium monitoring PDs with 5\% coupling ratio are used to set the PMI phase conditions. The output of the MZI is terminated to germanium BPDs, with responsivity of 0.96 A W$^{-1}$, through a 2$\times$2 MMI with insertion loss of 0.15 dB. The photonic chip dimensions are 4.8 mm $\times$ 0.84 mm.

\subsection*{Phase noise measurement}
A low phase noise reference laser at 1548.42 nm, with a sub-Hz Lorentzian linewidth, is PDH-locked to the fundamental mode of an ultra-stable Fabry–P\'erot cavity constructed with ULE spacers and held in a vacuum chamber. A fraction of the stabilized laser output serves as the frequency reference for locking an octave-spanning fiber frequency comb (FFC), thereby stabilizing its 200 MHz repetition rate. The FFC is self-referenced, and its carrier–envelope offset frequency ($f_{CEO}$) is stabilized with an integrated phase noise below 0.8 rad. The fully stabilized FFC output is amplified with an optical amplifier and filtered by an optical band-pass filter centered about the wavelength of the laser under test. A small portion of the stabilized laser output is tapped for out-of-loop frequency noise characterization. The output of the laser is mixed with the FFC and the generated beat note is measured using a phase noise analyzer and frequency counter for offset frequencies $<$ 1 kHz. A fiber based DSH setup is used to measure the laser frequency noise PSD for offset frequencies $>$ 1 kHz. The measured PSD is stitched to that of the FFC heterodyne beat note measurement to avoid fiber phase noise in DSH setup at low offset frequencies. The list of all equipment and tools are provided in Supplementary Table 1. 

\subsection*{Photonic chip packaging}
As shown in Fig. \ref{fig_pic}d, the PIC is mounted on a custom-designed, temperature-controlled aluminum submount. A custom-designed printed circuit board (PCB) is attached to the submount and wire-bonded to the chip. The laser is coupled into the PIC through a single-mode optical fiber that is glued to the input facet using UV-cured adhesive. On the output side, a GRIN lens with 1 mm diameter is attached to the PIC facet to match the mode size to that of the cavity, with a mode-field diameter of approximately 240~$\mu$m, thereby enabling efficient coupling with minimal loss. The GRIN lens is actively aligned to the PIC while monitoring the collimated output mode on an optical beam profiler.   

\subsection*{Cavity enclosure details}
The $\mu$FP cavity is sensitive to acoustic noise and air-pressure fluctuations, which induce random perturbations in its resonance frequency. To mitigate this, our experiments show that it is essential to package the cavity inside an airtight enclosure. For this purpose, a custom enclosure, with 2 mm thick walls surrounding the $\mu$FP cavity, was designed and 3D printed using commercially available Ceramic-Like Advanced High Temp White material, with two openings for the cavity input and output. Non-reflective transparent glass windows were affixed to these openings using UV-cured adhesive, and the enclosure lid was sealed with a rubber sheet to ensure airtightness.

\newpage
\section*{Data availability}
All data supporting the findings of this study are available within the
article and its supplementary file. Any additional requests for information can be directed to, and will be fulfilled by, the corresponding authors.

\section*{Acknowledgments}
We thank Nicolas K. Fontaine for helpful technical discussions. M.H.I, F.A, and K.K acknowledge support from Nokia Bell Labs. P.T.R, H.C, B.L, and N.J acknowledge support from Yale University, Quantum CT, and the Roberts Innovation Fund.

\section*{Author contributions}
M.H.I and H.C conceived the idea. M.H.I taped-out the photonic chip. H.C and N.J fabricated and characterized the $\mu$FP cavity. M.H.I and F.A conducted the experiments. M.H.I and H.C designed the PCB and package. B.L helped with the measurement and packaging design. K.K packaged the PIC. F.Q and P.T.R advised on the project. M.H.I wrote the paper with input from all authors. All authors reviewed the paper.  

\section*{Competing interests}
P.T.R. is a founder and shareholder of Resonance Micro Technologies Inc. The remaining authors declare no competing interests.

\section*{Additional information}
Correspondence and requests for materials should be addressed to Mohamad Hossein Idjadi or Peter T. Rakich.

\bibliography{ref.bib}

\end{document}


\title{\LARGE Chip-scale modulation-free laser stabilization using vacuum-gap micro-Fabry-P\'erot cavity}

\author[1,*]{\normalsize Mohamad Hossein Idjadi}
\author[2]{\normalsize Haotian Cheng}
\author[1]{\normalsize Farshid Ashtiani}
\author[2]{\normalsize Benjia Li}
\author[1]{\normalsize Kwangwoong Kim}
\author[2]{\normalsize Naijun Jin}
\author[3]{\normalsize Franklyn Quinlan}
\author[2,$\dagger$]{\normalsize Peter T. Rakich}
\affil[1]{\footnotesize Nokia Bell Labs, 600 Mountain Ave, Murray Hill, NJ 07974, USA.}
\affil[2]{\footnotesize Department of Applied Physics, Yale University, New Haven, CT 06511, USA}
\affil[3]{\footnotesize Time and Frequency Division, National Institute of Standards and Technology, Boulder, CO 80305, USA}
\affil[*, $\dagger$]{\footnotesize Corresponding authors: mohamad.idjadi@nokia-bell-labs.com, peter.rakich@yale.edu}

\begin{abstract}
\end{abstract}
\maketitle

\section*{Supplementary Note 1: On-chip Poor Man's isolator}
\begin{figure}[tb]
    \centering
    \includegraphics[width=\textwidth]{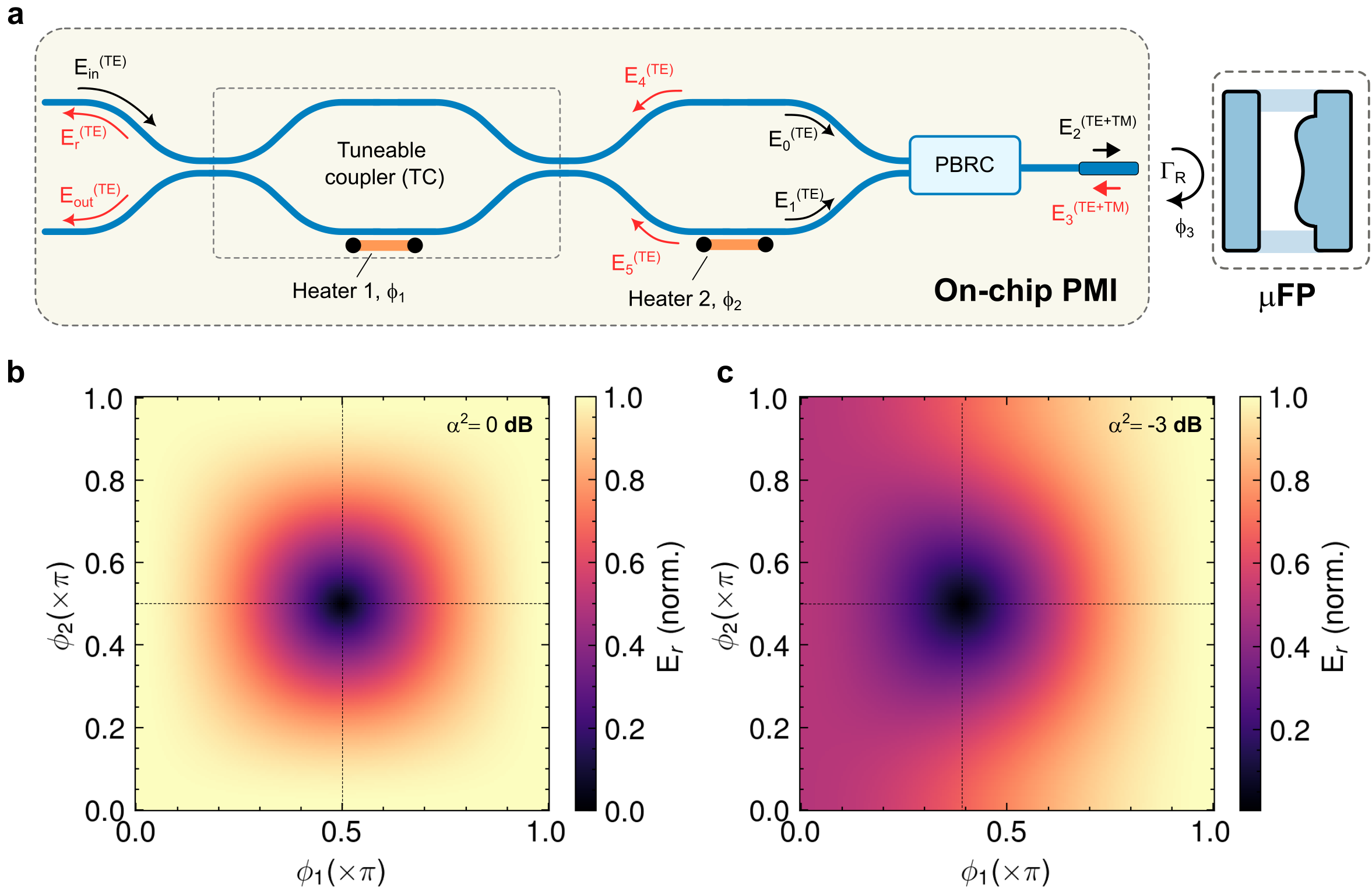}
    \caption{\textbf{| On-chip Poor Man's isolator (PMI)}. \textbf{a,} Architecture of the on-chip PMI. Simulated reflected E-field as a function of phases $\phi_1$ and $\phi_2$ for \textbf{b,} $\alpha^2 = 0$ dB and \textbf{c,} $\alpha^2 = -3$ dB power mismatch between the two modes. }
    \label{suppfig_pmi}
\end{figure}
Supplementary Figure \ref{suppfig_pmi}a shows the structure of the PMI, based on our previous work \cite{cheng2023novel}, with the input TE-mode electric field (E-field) denoted as $E_{in}^{(TE)}$. 

A tunable coupler (TC), implemented as a Mach-Zehnder interferometer (MZI), divides the input signal into two fields with a $\pi$ radian phase difference. The transfer matrix of a TC is 
\begin{equation}
\label{eq1}
        M_{TC}(\phi_1)=\frac{1}{2} 
    \begin{pmatrix}
    1-e^{-j\phi_1} & -j(1+e^{-j\phi_1})\\
    -j(1+e^{-j\phi_1}) & -(1-e^{-j\phi_1})
    \end{pmatrix}.
\end{equation}

Using Supplementary Eq. (\ref{eq1}), the output of the TC can be written as 
\begin{equation}
    \begin{pmatrix}
        E_0^{(TE)}\\
        E_1^{(TE)}
    \end{pmatrix}
    =
    \begin{pmatrix}
        1 & 0\\
        0 & e^{-j\phi_2}
    \end{pmatrix}
    \times
    M_{TC}(\phi_1)
    \times
    \begin{pmatrix}
        E_{in}^{(TE)}\\
        0
    \end{pmatrix},
\end{equation}
where $\phi_1$ and $\phi_2$ are the optical phases controlled by thermal phase shifters. When $\phi_1=\frac{\pi}{2}$, 
\begin{equation}
    \label{eq4}
    \begin{pmatrix}
    E_0^{(TE)}\\
    E_1^{(TE)}
    \end{pmatrix}
    =\frac{e^{j\pi/4}}{\sqrt{2}}E_{in}
    \begin{pmatrix}
    1\\
    -e^{-j\phi_2}
    \end{pmatrix}.
\end{equation}
\\
\\
A polarization beam rotator-combiner (PBRC) is a three-port device with two inputs; it rotates the polarization of one input and combines it with the other at the output:
\begin{gather}
    E_2^{(TE+TM)}=E_0^{(TE)}+\alpha E_1^{(TM)}, \label{eq5}
\end{gather}
where $\alpha^2$ is the optical power mismatch between TE and TM modes. Ideally, we can assume $\alpha =1$. 
The Jones matrix of the cavity is 
$\begin{pmatrix}
\Gamma_R(\omega) & 0\\
0 & \Gamma_R(\omega)
\end{pmatrix}$,
hence the reflected E-field from the micro Fabry-P\'erot ($\mu$FP) cavity is
\begin{gather}
    E_3^{(TE+TM)}=\Gamma_R(\omega) \big(E_0^{(TE)}+E_1^{(TM)} \big)e^{-j\phi_3}, \label{eq6}
\end{gather}
where $\Gamma_R(\omega)$ is the E-field reflection coefficient of the cavity and $\phi_3$ is the round-trip phase accumulated in free-space. The reflected field couples back into the PBRC and divides into two TE modes: 
\begin{gather}
    E_4^{(TE)}=\frac{\Gamma_R(\omega)e^{-j\phi_3}}{\sqrt{2}}e^{+j\pi/4} E_{in}^{(TE)}, \label{eq7}\\
    E_5^{(TE)}=\frac{-\Gamma_R(\omega)e^{-j\phi_3}}{\sqrt{2}}e^{+j\pi/4} E_{in}^{(TE)}e^{-j2\phi_2}. \label{eq8}
\end{gather}

Using Supplementary Eqs. (\ref{eq1}), (\ref{eq7}), and (\ref{eq8}), the reflection and transmission fields can be simplified to
\begin{equation}
\label{eq10}
    \begin{pmatrix}
    E_r^{(TE)}\\
    E_{out}^{(TE)}
    \end{pmatrix}
    =\frac{j\Gamma_R(\omega)e^{-j\phi_3}}{2} E_{in}^{(TE)}
    \begin{pmatrix}
    1+e^{-j2\phi_2}\\
    -1+e^{-j2\phi_2}
    \end{pmatrix}.
\end{equation}

If $\phi_2=\frac{\pi}{2}$, the reflected beam is canceled 
\begin{equation}
\label{eq11}
    \begin{pmatrix}
    E_r^{(TE)}\\
    E_{out}^{(TE)}
    \end{pmatrix}
    =E_{in}^{(TE)}
    \begin{pmatrix}
    0\\
    -j\Gamma_R(\omega)e^{-j\phi_3}
    \end{pmatrix}.
\end{equation}
As shown in Supplementary Fig. \ref{suppfig_pmi}b, if the TE and TM polarizations match in amplitude, setting $\phi_1=\phi_2=\frac{\pi}{2}$ results in complete cancellation of the reflected field which is in agreement with Supplementary Eq. (\ref{eq11}). The calculations above assume an ideal scenario where the TE and TM polarizations launched from the PBRC have equal amplitudes. The fields
$\begin{pmatrix}
    E_r\\
    E_{out}
\end{pmatrix}$
can also be calculated when there is an amplitude mismatch factor $\alpha$.  As shown in Supplementary Fig. \ref{suppfig_pmi}c, with an optical power mismatch (e.g. $\alpha^2=-3$ dB), the PMI circuit can still null the reflected field by adjusting $\phi_1$ via the thermal phase shifter.

\newpage
\section*{Supplementary Note 2: CCMZI error signal using $\mu$FP cavity}
\begin{figure}[b!]
    \centering
    \includegraphics[width=\textwidth]{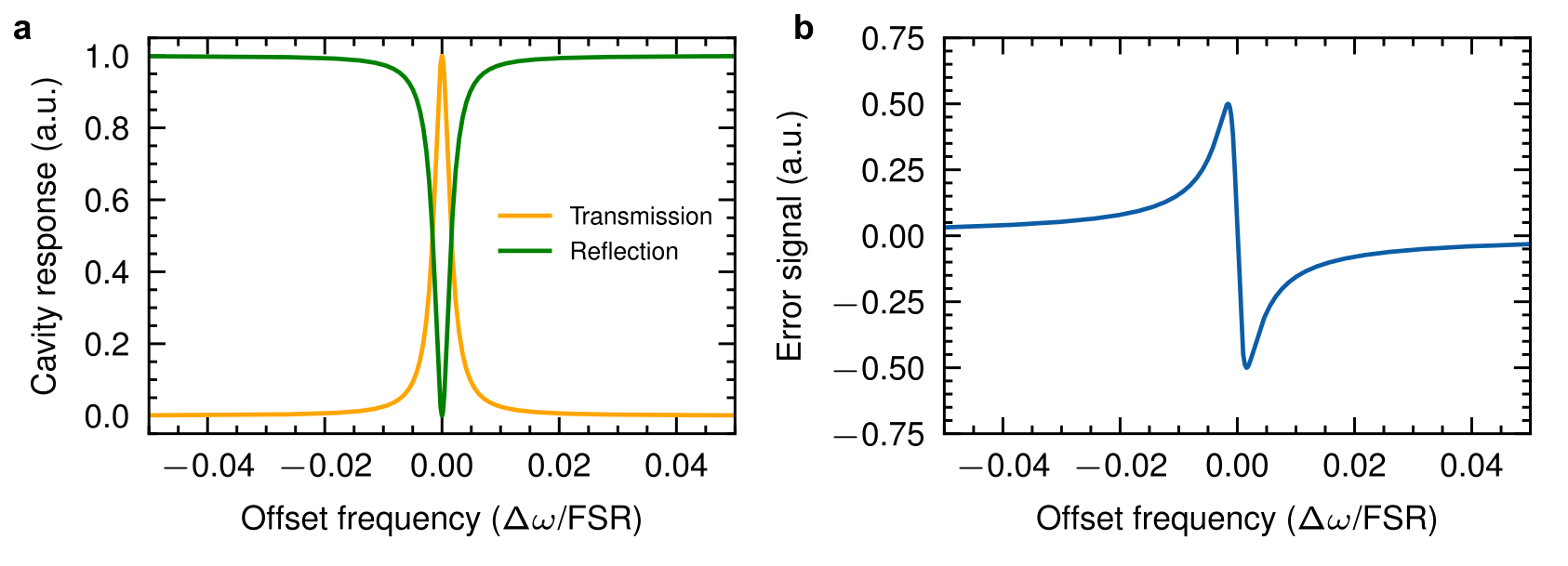}
    \caption{\textbf{| CCMZI error signal simulation.} \textbf{a,} Simulated reflection and transmission spectra of a Fabry–P\'erot cavity with 99\% mirror reflectivity. \textbf{b,} Corresponding error signal obtained when the reflected light in (\textbf{a}) is utilized by the CCMZI.}
    \label{suppfig_error}
\end{figure}
A FP cavity consists of two optically reflective surfaces, i.e. mirrors, with E-field reflectivity of $\rho$. As a result, the reflection and transmission coefficients for E-field can be written as
\begin{gather}
   \Gamma_R(\varphi)=\rho \frac{e^{-j\varphi}-1}{1-\rho^2e^{-j\varphi}},\\ 
   \Gamma_T(\varphi)=(1-\rho^2)\frac{e^{-j\frac{\varphi}{2}}}{1-\rho^2e^{-j\varphi}}, 
\end{gather}
where $\varphi, \Gamma_T(\varphi)$, and $\Gamma_R(\varphi)$ are round trip phase of the light traveling inside the cavity, amplitude transmission coefficient and amplitude reflection coefficient for the E-field, respectively. As shown by Eq. (1), since the $\mu$FP cavity is used in reflection mode in the CCMZI circuit, the transfer function of the cavity is $\Gamma_R(.)$ and, hence, the generated error signal is \cite{idjadi2024modulation}
\begin{equation}
    i_{error}(\omega)=RP_0|\Gamma_R(\omega)|\times \mathrm{sin}(\angle \Gamma_R(\omega)-\phi_0), 
\end{equation}
where $i_{error}, R, P_0, \angle \Gamma_R(.), |\Gamma_R(.)|, \phi_0$ are the error signal, responsivity of the BPDs, intensity of the laser at the input of the MZI, phase and magnitude of the transfer function, and static phase introduced by the thermal phase shifter in the MZI, respectively. 
Supplementary Figure \ref{suppfig_error}a shows the simulated transmission and reflection response of a FP cavity (with $\rho^2=0.99$) as a function of offset from the cavity resonance frequency. Supplementary Figure \ref{suppfig_error}b shows the corresponding normalized error signal with an asymmetric response around the FP cavity resonance which matches the measured open loop response in Fig. 2c.

\newpage
\section*{Supplementary Note 3: $\mu$FP cavity ring-down characterization}
\begin{figure}[h]
    \centering
    \includegraphics[width=\textwidth]{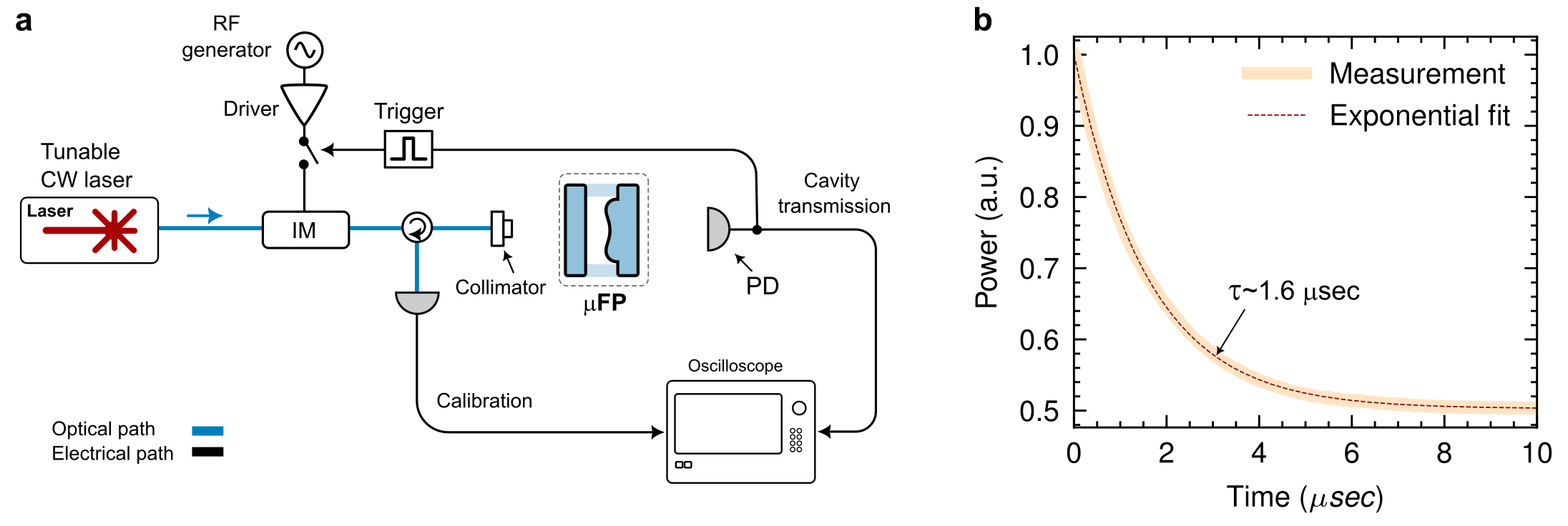}
    \caption{\textbf{| Ring-down characterization of the $\mu$FP cavity.} \textbf{a,} Measurement setup for cavity ring-down experiment. \textbf{b,} Optical ring-down of the $\mu$FP cavity shows a photon decay time of about 1.6 $\mu$sec. IM intensity modulator, PD photodetector.}
    \label{fig_supp_ringdwn}
\end{figure}

Supplementary Figure \ref{fig_supp_ringdwn}a shows the cavity ring-down measurement setup where a tunable CW laser, at wavelength of 1555 nm, is scanned across the cavity resonance and the transmission signal triggers the RF switch and shuts off the laser intensity through an intensity modulator\cite{jin2022micro}. Supplementary Figure \ref{fig_supp_ringdwn}b shows the ring-down measurement with measured photon decay time of about 1.6 $\mu$sec which corresponds to a Q-factor of about $2\times10^9$.
\\

\section*{Supplementary Table}

\begin{table}[htb]
    \small
	\centering
  \begin{tabular}{|l|l|}
  	\hline
    \textbf{Equipment} & \textbf{Model} \\
    \hline
    Fiber laser			        &		NKT-E15                          \\
    Reference laser             &       OEwaves OE4040-XLN              \\
    Frequency comb              &       Vescent FFC-100                  \\ 
    Photodetector               &       PDB450C                          \\
    Temperature controller		&		Wavelength Electronics-TC10      \\		
    Oscilloscope				&		RIGOL DS1104	                  \\
    Power supply				&		SIGLENT SPD3303X-E	            \\
    PID controller              &       Vescent D2-125                  \\
    Optical filter              &       Santec OTF-350                  \\
    AOM                         &       CSRAYZER-AOM-1550-200M          \\
    Optical spectrum analyzer	&		YOKOGAWA AQ6370D			    \\
    RF spectrum analyzer        &       Agilent 89441A                  \\
    Phase noise analyzer        &       Microchip 53100A                \\
    TIA                         &       Femto DHPCA-100                 \\
    AOM driver                  &       Mini-circuits ZHL-1A-S+         \\ 
    VCO                         &       Mini-circuits ZOS-300+          \\
    \hline
  \end{tabular}
  \caption{| List of equipment and components used in different experimental setups. }
\end{table}
\small{
\bibliography{supp_ref.bib}
}